\documentstyle[12pt]{article}
\newcommand\be{\begin{equation}}
\newcommand\ee{\end{equation}}
\newcommand\bea{\begin{eqnarray}}
\newcommand\eea{\end{eqnarray}}
\newcommand\ket[1]{|#1\rangle}

\newcommand\braket[2]{\langle #1|#2\rangle}
\newcommand{\fatalpha}{{\bf \alpha \kern -0.44em \alpha}}
\newcommand{\fatsigma}{{\bf \sigma \kern -0.54em \sigma}}
\newcommand{\tpchi}{{\bf \chi \kern -0.35em \chi}}
\newcommand{\llambda}{{\bf \lambda \kern -0.45em \lambda}}

\renewcommand{\theequation}{\arabic{equation}}
\renewcommand{\theequation}{\thesection-\arabic{equation}}
\bibliography{plain}
\title{\bf Quantum central limit theorem for continuous-time quantum walks on
odd graphs in quantum probability theory}\vspace{20mm}
\author{ S. Salimi$^{a}$
  \thanks{E-mail:shsalimi@uok.ac.ir}
 \\ $^a${\small Department of Physics,
University of Kurdistan, Kurdistan 51664, Iran.} }  \pagebreak

\vspace{20mm}
\begin{document}
\maketitle \vspace{15mm}
\newpage
\begin{abstract}

The method of the quantum probability theory only requires simple
structural data of graph and allows us to avoid a heavy
combinational argument often necessary to obtain full description of
spectrum of the adjacency matrix.  In the present paper, by using
the idea of calculation of the probability amplitudes for
continuous-time quantum walk in terms of the quantum probability
theory, we investigate quantum central limit theorem for
continuous-time quantum walks on odd graphs.


{\bf PACs Index: 03.65.Ud }
\end{abstract}

\vspace{70mm}
\newpage
\section{Introduction}
Two types of quantum walks, discrete  and continuous time, were
introduced as the quantum mechanical extension of the corresponding
random walks and have been extensively studied over the last few
years \cite{adz, Kempe}.

Random walks on graphs are the basis of a number of classical
algorithms. Examples include 2-SAT  (satisfiability for certain
types of Boolean formulas), graph connectivity, and finding
satisfying assignments for Boolean formulas.  It is this success of
random walks that motivated the study of their quantum analogs in
order to explore whether they might extend the set of quantum
algorithms. Several systems have been proposed as candidates to
implement quantum random walks. These proposals include atoms
trapped in optical lattices \cite{ap1}, cavity quantum
electrodynamics (CQED) \cite{ap2} and nuclear magnetic resonance
(NMR) in solid substrates \cite{ap3,ap4}. In liquid-state NMR
systems \cite{ap5}, time–resolved observations of spin waves has
been done \cite{ap6}. It has also been pointed out that a quantum
walk can be simulated using classical waves instead of matter waves
\cite{ap7,ap8}.

A study of quantum walks on simple graph is well known in
physics(see \cite{fls}). Recent studies of quantum walks on more
general graphs were described in \cite{ccdfgs,
fg,cfg,abnvw,aakv,mr,kem}. Some of these works studies the problem
in the important context of algorithmic problems on graphs and
suggests that quantum walks is a promising algorithmic technique for
designing future quantum algorithms.

There is one approach for investigation of continuous-time quantum
walk (CTQW) on graphs  using the spectral distribution associated
with the adjacency matrix of graphs \cite{js,jsa, jsas, konno1,
konno2}. Authors in Ref.\cite{js} have introduced a new method for
calculating the probability amplitudes of quantum walk based on
spectral distribution which allows us to avoid a heavy combinational
argument often necessary to obtain full description of spectrum of
the Hamiltonian. It is interesting to investigate  the CTQW on graph
when the graph grows as time goes by. In order to study a quantum
system in full detail, its Hamiltonian needs to be diagonalized.
With increasing dimension of the Hilbert space, the diagonalization
of an operator becomes a very tedious task. In fact, we discuss this
question in the CTQW as a quantum central limit theorem for CTQW. In
this paper we try to investigate quantum central limit theorem for
CTQW on growing odd graph via spectral distribution.

The organization of the paper is as follows. In section 2, we give a
brief outline of  graphs and introduce odd graph. In Section $3$, we
review the stratification and quantum decomposition for adjacency
matrix of graphs. Section $4$ is devoted to the method of computing
the amplitude for CTQW, through spectral distribution $\mu$ of the
adjacency matrix. In the first subsection of section $5$ we evaluate
the CTQW on finite odd graph and the in second subsection $5$, we
investigate quantum central limit theorem for CTQW on odd graph. The
paper is ended with a brief conclusion and one appendix which
contains determination of spectral distribution by continued
fractions method.

\section{Odd graph}
In beginning we present a summery of graph and then we introduce odd
graph.

A graph is a pair $G=(V,E)$, where $V$ is a non-empty set and $E$ is
a subset of $\{\{\alpha,\beta\}|\alpha, \beta\in V,\alpha\neq \beta
\}$. Elements of $V$ and  $E$ are called \emph{vertices} and
\emph{edges}, respectively. Two vertices $\alpha, \beta\in V$ are
called adjacent if $\{\alpha,\beta\}\in E$, and in this case we
write $\alpha\sim \beta$. Let $l^2(V)$ denotes the Hilbert space of
C-valued square-summable functions on
 V, and $\{\ket{\alpha}|\ \alpha\in V\}$ becomes a complete orthonormal basis of
$l^2(V)$. The adjacency matrix $A=(A_{\alpha
\beta})_{\alpha,\beta\in V}$ is defined by

\[
A_{\alpha \beta} = \left\{
\begin{array}{ll}
1 & \mbox{if $ \alpha\sim \beta$}\\
0 & \mbox{otherwise.}
\end{array}
\right.
\]

which is considered as an operator acting on $l^2(V)$ in such a way
that
$$
A\ket{\alpha}=\sum_{\alpha\sim \beta}\ket{\beta}, \;\;\;\;\alpha\in
V.
$$
Obviously, (i) $A$ is symmetric; (ii) an element of $A$ takes a
value in $\{0, 1\}$; (iii)  a diagonal element of $A$ vanishes.
Conversely, for a non-empty set $V$, a graph structure is uniquely
determined by such a matrix indexed by $V$. The \emph{degree} or
\emph{valency} of a vertex $\alpha\in V$ is defined by
$$
\kappa(\alpha)=|\{\beta\in V| \alpha\sim \beta\}|,
$$
where $|.|$ denotes the cardinality.

Let $S$ be a set of integer as $S=\{1,2,...,2k-1\}$ for a fixed
integer $k\geq 2$. Now we define $V$ be the set of subsets of $S$
having cardinality $k-1$, i.e.,
\begin{equation}
V=\{\alpha \subset S\ |\alpha|=k-1\},
\end{equation}
and put
\begin{equation}
E=\{\{\alpha,\beta\}|\alpha, \beta\in V,\alpha\cap \beta=\emptyset
\}.
\end{equation}
This graph $(V,E)$ is called the odd graph of degree $k$  and is
denoted by $O_k$. Obviously, $O_k$ is a regular graph with degree
$k$. For $k=3$, the odd graph is  well-known as Petersen graph
(Fig.1). The odd graphs have been studied in algebraic graph theory
where  some of their properties are found in
Refs.\cite{biggs1,biggs2}.

For $n=0,1,2,...,k-1$ we define $\epsilon_n$ as
\[
\epsilon_n = \left\{
\begin{array}{ll}
k-1-\frac{n}{2} & \mbox{if $n$ is even}\\
\frac{n-1}{2} & \mbox{if $n$ is odd,}
\end{array}
\right.
\]
then by using proposition $4.1$ \cite{nob1}, for pair $\alpha,
\beta\in V$ we have
\begin{equation}
 \;\;\;\ |\alpha\cap \beta|=\epsilon_n \;\;\;\
\Longleftrightarrow
\partial(\alpha, \beta)=n,
\end{equation}
where $\partial$  stands for the natural distance function. Due to
the above relations based on distance function the odd graphs are
distance regular graphs, i.e., for a given $i,j,l=0,1,2,...,$ the
intersection number
\begin{equation}
p_{ij}^l=|\{\gamma\in V| \partial(\alpha, \gamma)=i \ \mbox{and}
\;\;\
\partial(\gamma, \beta)=j\}|,
\end{equation}
is independently determined  of the choice of $\alpha, \beta\in V$
satisfying $\partial(\alpha, \beta)=l$. There are some well-known
facts about the intersection numbers of distance regular graphs, for
example $p_{j1}^i=0 $ (for $i\neq 0$, $j$ is not $\{i-1, i, i+1
\}$)( for more details see Refs. \cite{cohen, bailey}). For
convenience, set $b_i:=p_{i-1,1}^i (1\leq i\leq d)$,
$c_i:=p_{i+1,1}^i (0\leq i \leq d-1)$, $a_i:=p_{i,1}^i (0\leq i \leq
d)$, $k_i:=p_{i,i}^0 (0\leq i \leq d)$ and $b_0=c_d=0$, where
$d:=$max$\{\partial(\alpha, \beta): \alpha, \beta\in V \}$ is called
diameter of graph. Moreover
\begin{equation}
b_i+a_i+c_i=k, \;\;\;\;\ 0\leq i \leq d,
\end{equation}
where $k=k_1$ is degree of graph.

For calculation of  Szeg\"{o}- Jacobi sequences $\{\omega_k\}$ and
$\{\alpha_k\}$  in the next section we will require $b_i$ and $c_i$
\cite{jsa}, where for odd graph is given by

\begin{equation}\label{interodd1}
b_i\;=\;\cases{\frac{i}{2} & if $\;i \; \mbox{is even} $,\cr
\frac{i+1}{2} & if $\;i \; \mbox{is odd} $\cr}\qquad \qquad (1\leq
i\leq k-1),
\end{equation}
\begin{equation}\label{interodd2}
c_i\;=\;\cases{k-\frac{i}{2} & if $\;i \; \mbox{is even} $,\cr
k-\frac{i+1}{2} & if $\;i \; \mbox{is odd} $\cr}\qquad \qquad (0\leq
i\leq k-2).
\end{equation}

\section{Quantum Probabilistic Approach for CTQW}
In this section we give some preliminaries that require to describe
CTQW via spectral distribution technique.

\subsection{Stratification}
Due to definition of function $\partial$, the graph becomes a metric
space with the distance $\partial$. We fix a point $o\in V$ as an
origin of the graph, called reference vertex. Then, the graph is
stratified into a disjoint union of strata:
\begin{equation}\label{v1}
V=\bigcup_{i=0}^{\infty}V_i,\;\;\;\;\;\; V_i=\{\alpha\in V|\
\partial(o,\alpha)=i\}.
\end{equation}
With each stratum $V_i$ we associate a unit vector in $l^2(V)$
defined by
\begin{equation}
\ket{\phi_{i}}=\frac{1}{\sqrt{|V_i|}}\sum_{\alpha\in V_{i}}\ket{i,
\alpha},
\end{equation}
where $\ket{i, \alpha}$ denotes the eigenket of the $\alpha$th
vertex at the stratum $i$.
 The closed subspace of $l^2(V)$ spanned by
$\{\ket{\phi_{i}}\}$ is denoted by $\Lambda(G)$. Since
$\{\ket{\phi_{i}}\}$ becomes a complete orthonormal basis of
$\Lambda(G)$, we often write
\begin{equation}
\Lambda(G)=\sum_{i}\oplus \textbf{C}\ket{\phi_{i}}.
\end{equation}

\subsection{Quantum decomposition}
One can obtain a quantum decomposition associated with the
stratification (\ref{v1}) for the adjacency matrices of this type of
graphs as
\begin{equation}
A=A^{+}+A^{-}+A^0.
\end{equation}
where three matrices $A^+$, $ A^-$ and $A^0$  are defined as
follows: for $\alpha\in V_i$
\[
(A^+)_{\beta\alpha} = \left\{
\begin{array}{ll}
A_{\beta\alpha} & \mbox{if $ \beta\in V_{i+1}$}\\
0 & \mbox{otherwise,}
\end{array}
\right.
\]
\[
(A^-)_{\beta\alpha} = \left\{
\begin{array}{ll}
A_{\beta\alpha} & \mbox{if $ \beta\in V_{i-1}$}\\
0 & \mbox{otherwise,}
\end{array}
\right.
\]
\[
(A^0)_{\beta\alpha} = \left\{
\begin{array}{ll}
A_{\beta\alpha} & \mbox{if $ \beta\in V_{i}$}\\
0 & \mbox{otherwise,}
\end{array}
\right.
\]
or, equivalently, for $\ket{i,\alpha}$,
\begin{equation}\label{qd}
A^{+}\ket{i,\alpha}=\sum_{\beta\in V_{i+1}}\ket{i+1,\beta},
\;\;\;\;\ A^{-}\ket {i,\alpha}=\sum_{\beta\in
V_{i-1}}\ket{i-1,\beta}, \;\;\;\;\
A^{0}\ket{i,\alpha}=\sum_{\beta\in V_{i}}\ket{i,\beta}, \;\;\;\;\
 \end{equation}
 for $\{\alpha, \beta\}\in E$.
Since $\alpha\in V_i$ and $\{\alpha,\beta\}\in E$ then $\beta\in
V_{i-1}\bigcup V_i\bigcup V_{i+1}$, where we tacitly understand that
$V_{-1}=\emptyset$.  The vector state corresponding to
$\ket{o}=\ket{\phi_0}$, with $o\in V$ as the fixed origin, is
analogous to the vacuum state in Fock space.
 According to Ref.\cite{nob},
$<A^m>$ coincides with the number of $m$-step walks starting and
terminating at $o$, also, by lemma 2.2, \cite{nob} if $G$ is
invariant under the quantum components $A^\varepsilon$,
$\varepsilon\in \{+,-,0\}$,  then there exist two Szeg\"{o}- Jacobi
sequences $\{\omega_i\}_{i=1}^{\infty}$ and
$\{\alpha_i\}_{i=1}^{\infty}$ derived from $A$, such that
\begin{equation}\label{v5}
A^{+}\ket{\phi_{i}}=\sqrt{\omega_{i+1}}\ket{\phi_{i+1}}, \;\;\;\
i\geq 0
\end{equation}
\begin{equation}\label{v6}
A^{-}\ket{\phi_{0}}=0, \;\;\
A^{-}\ket{\phi_{i}}=\sqrt{\omega_{i}}\ket{\phi_{i-1}}, \;\;\;\ i\geq
1
\end{equation}
\begin{equation}\label{v7}
A^{0}\ket{\phi_{i}}=\alpha_{i+1}\ket{\phi_{i}}, \;\;\;\ i\geq 0,
\end{equation}
where
$\sqrt{\omega_{i+1}}=\frac{|V_{i+1}|^{1/2}}{|V_{i}|^{1/2}}\kappa_{-(\beta)}$,
$\kappa_{-(\beta)}=|\{\alpha\in V_i| \alpha\sim \beta\}|$ for
$\beta\in V_{i+1}$ and $\alpha_{i+1}=\kappa_{0(\beta)}$, such that
$\kappa_{0(\beta)}=|\{\alpha\in V_i| \alpha\sim \beta\}|$ for
$\beta\in V_i$. In particular $(\Lambda(G), A^+, A^-)$ is an
interacting Fock space associated with a Szeg\"{o}-Jacobi sequence
$\{\omega_i\}$.

\subsection{Study  of CTQW on a graph via spectral distribution of
its adjacency matrix} The CTQW  on graph has been introduced as the
quantum mechanical analogue of its classical counterpart, which is
defined by replacing Kolmogorov's equation (master equation) of
continuous-time classical random walk on a graph
\cite{ghwiss,nvkampen}
\begin{equation}
\frac{dP_m(t)}{dt} = \sum_{n=1}^{l}H_{mn}P_j(t), \;\;\ m=1,2,...,l
\end{equation}
with Schr\"{o}dinger's equation. Matrix $H$ is the Hamiltonian of
the walk and $P_m(t)$ is the occupying probability of vertex $m$ at
time $t$. It is natural to choose the Laplacian of the graph,
defined as $L=A-D$ as the Hamiltonian of the walk, where $D$ is a
diagonal matrix with entries $D_{jj}=deg(\alpha_j)$.

Let $\ket{\phi(t)}$ be a time-dependent amplitude of the quantum
process on  graph $\Gamma$. The wave evolution of the quantum walk
is
\begin{equation}
 i\hbar\frac{d}{dt}\ket{\phi(t)} = H\ket{\phi(t)},
\end{equation}
where from now on  we assume $\hbar = 1$,  and $\ket{\phi_{0}}$ is
the initial amplitude wave function of the particle. The solution is
given by $\ket{\phi(t)} = e^{-iHt} \ket{\phi_{0}}$.  On $d$-regular
graphs, $D = \frac{1}{d}I$, and since $A$ and $D$ commute, we get
\begin{equation} \label{eqn:phase-factor}
e^{-itH} = e^{-it(A-\frac{1}{d}I)} = e^{-it/d}e^{-itA}.
\end{equation}
This introduces an irrelevant phase factor in the wave evolution.
Hence we can consider $H=A$. Thus, we have
\begin{equation}
 \ket{\phi(t)} = e^{-iAt}\ket{\phi_0},
\end{equation}

One of our goals in this paper is the evaluation of probability
amplitudes for CTQW  on graphs by using the method of spectral
distribution associated with the adjacency matrix. The spectral
properties of the adjacency matrix of a graph play an important role
in many branches of mathematics and physics. The spectral
distribution can be generalized in various ways. In this work,
following Refs.\cite{js, nob}, we consider the spectral distribution
$\mu$ of the adjacency matrix $A$:
\begin{equation}\label{v2}
\langle A^m\rangle=\int_{R}x^{m}\mu(dx), \;\;\;\;\ m=0,1,2,...
\end{equation}
where $\langle.\rangle$ is the mean value with respect to the state
$\ket{\phi_0}$. By condition  of quantum decomposition (QD) graphs
the ''moment'' sequence $\{\langle A^m\rangle\}_{m=0}^{\infty}$ is
well-defined\cite{js,nob}. Then the existence of a spectral
distribution satisfying (\ref{v2}) is a consequence of Hamburger's
theorem, see e.g., Shohat and Tamarkin [\cite{st}, Theorem 1.2].

Due to using the quantum decomposition relations (\ref{v5},
\ref{v6},\ref{v7}) and the recursion relation (\ref{op}) of
polynomial $P_n(x)$, the other matrix elements $\label{cw1}
\braket{\phi_{n}}{A^m\mid \phi_0}$ can be written as
\begin{equation}\label{cw1}
\braket{\phi_{n}}{A^m\mid
\phi_0}=\frac{1}{\sqrt{\omega_1\omega_2\cdots \omega_{n}
}}\int_{R}x^{m}P_{n}(x)\mu(dx),  \;\;\;\;\ m=0,1,2,....
\end{equation}
where  is useful for obtaining of amplitudes of CTQW in terms of
spectral distribution associated with the adjacency matrix of graphs
\cite{js}.

Therefore by using (\ref{cw1}), the probability amplitude of
observing the walk at stratum $m$ at time $t$ can be obtained as
\begin{equation}\label{v4}
q_{m}(t)=\braket{\phi_{m}}{\phi(t)}=\braket{\phi_{m}}{e^{-iAt}\mid
\phi_0}=\frac{1}{\sqrt{\omega_1\omega_2\cdots\omega_{m}}}\int_{R}e^{-ixt}P_{m}(x)\mu(dx).
\end{equation}
The conservation of probability $\sum_{m=0}{\mid
\braket{\phi_{m}}{\phi(t)}\mid}^2=1$ follows immediately from
Eq.(\ref{v4}) by using the completeness relation of orthogonal
polynomials $P_n(x)$. In the appendix $A$ of the reference
\cite{js}, it is proved that the walker has the same amplitude at
the vertices belonging to the same stratum, i.e., we have
$q_{im}(t)=\frac{q_{m}(t)}{\mid V_m\mid}, i=0,1,...,\mid V_m\mid$,
where $q_{im}(t)$ denotes the amplitude of the walker at $i$th
vertex of $m$th stratum.

Investigation of CTQW via spectral distribution method  pave the way
to calculate CTQW on infinite graphs and to approximate with finite
graphs and vice versa, simply via Gauss quadrature formula, where in
cases of infinite graphs, one can study asymptotic behavior of walk
at large enough times by using the method of stationary phase
approximation (for more details see [1]).

Indeed, the determination of $\mu(x)$ is the main problem in the
spectral theory of operators, where this is quite possible by using
the continued fractions method, as it is explained in appendix A.

\section{Quantum central limit theorem for CTQW on odd graphs}
Having studied CTQW on finite odd graphs using the  method of the
spectral distribution, we investigate quantum central limit theorem
for CTQW on this graphs which is our main goals.

To consider stratification and quantum decomposition  of section
$3$,
 Ref.\cite{jsa} (i.e., $\omega_i=c_{i-1}b_{i}, \;\;\;\ \alpha_i=a_1-b_{i-1}-c_{i-1}$) and Eqs.(\ref{interodd1}),
 (\ref{interodd2}), we obtain two Szeg\"{o}- Jacobi sequences $\{\omega_i\}$
and $\{\alpha_i\}$ for $O_k$  as follows:\\
if $i$ is odd,
\begin{equation}\label{central2}
\omega_i=\frac{i+1}{2}(k-\frac{i-1}{2}),
\end{equation}
if $i$ is  even,
\begin{equation}
\omega_i=\frac{i}{2}(k-\frac{i}{2})
\end{equation}
if $i$ is odd or even,
\begin{equation}
\alpha_i=0.
\end{equation}
\textbf{A. Finite $k$ case}

 Let $\mu_k$ denote the spectral distribution of odd graph
$O_k$. Here for studying CTQW on finite odd graph we consider $k=4$
case. Then we have
\begin{equation}
\omega_1=4,\;\ \omega_2=3, \;\ \omega_3=6, \;\
\alpha_1=\alpha_2=\cdots= 0.
\end{equation}
Therefore we obtain Stieltjes transform
\begin{equation}
G_{\mu_4}(z)=\frac{z^3-9z}{z^4-13z^2+24}.
\end{equation}
In this case one can obtain the spectral distribution as follows
$$
\mu_4=\frac{\sqrt{73}}{292}(5+\sqrt{73})(\delta(x-\frac{1}{2}\sqrt{26-2\sqrt{73}})+
\delta(x+\frac{1}{2}\sqrt{26-2\sqrt{73}}))+
$$
\begin{equation}
\frac{\sqrt{73}}{292}(-5+\sqrt{73})(\delta(x-\frac{1}{2}\sqrt{26+2\sqrt{73}})+
\delta(x+\frac{1}{2}\sqrt{26+2\sqrt{73}}))
\end{equation}
By using Eq.(\ref{v4}) the amplitudes for walk at time $t$ are
$$
q_o(t)=\frac{\sqrt{73}}{146}((5+\sqrt{73})\cos
(\frac{1}{2}\sqrt{26-2\sqrt{73}})t+(-5+\sqrt{73})\cos(
\frac{1}{2}\sqrt{26+2\sqrt{73}})t),
$$
$$
q_1(t)=\frac{-i\sqrt{73}}{584}((5+\sqrt{73})(\sqrt{26-2\sqrt{73}})\sin
(\frac{1}{2}\sqrt{26-2\sqrt{73}})t+
$$$$
(-5+\sqrt{73})(\sqrt{26+2\sqrt{73}})\sin(
\frac{1}{2}\sqrt{26+2\sqrt{73}})t),
$$
$$
q_2(t)=\frac{2\sqrt{3}}{\sqrt{73}}(-\cos
(\frac{1}{2}\sqrt{26-2\sqrt{73}})t+ \cos(
\frac{1}{2}\sqrt{26+2\sqrt{73}})t),
$$
$$
q_3(t)=\frac{-i\sqrt{73}}{584\sqrt{2}}(-(13+\sqrt{73})(\sqrt{26-2\sqrt{73}})\sin
(\frac{1}{2}\sqrt{26-2\sqrt{73}})t+
$$
\begin{equation}
(-13 +\sqrt{73})(\sqrt{26+2\sqrt{73}})\sin(
\frac{1}{2}\sqrt{26+2\sqrt{73}})t).
\end{equation}
\textbf{B. Quantum central limit theorem }

In the limit of large $k\longrightarrow \infty$, it is observed  the
odd graphs $O_k$ form a growing family of distance regular graphs.
In the remaining of this section, we obtain CTQW on odd graphs $O_k$
for $k\longrightarrow \infty$ by applying the quantum central limit
theorem where is derived from the quantum probabilistic techniques.

\textbf{Theorem  (\cite{nob1})}. Let $\{G^{k}=(V^k, E^k)\}$ be a
growing family of  distance regular graphs. Let $A_k$ and
$\{p_{ij}^{l}(k)\}$ be the adjacency matrix and the intersection
numbers of $G^{k}$, respectively. Assume that the limits
$$
\omega_i=\lim_{k\longrightarrow\infty}\frac{p_{i-1,1}^{i}(k)p_{i,1}^{i-1}(k)}{p_{11}^{0}(k)}=\lim_{k\longrightarrow\infty}\frac{b_{i}(k)c_{i-1}(k)}{k}
$$
\begin{equation}\label{central1}
\alpha_i=\lim_{k\longrightarrow\infty}\frac{p_{i-1,1}^{i-1}(k)}{\sqrt{p_{11}^{0}(k)}}=\lim_{k\longrightarrow\infty}\frac{a_1(k)-b_{i-1}(k)-c_{i-1}(k)}{\sqrt{k}}
\end{equation}
exist for all $i=1,2,...$. Let $\Lambda=(\Lambda, \{\ket{\psi_i}\},
B^+, B^-)$ be the interacting Fock space associated with
$\{\omega_i\}$ and define a diagonal operator $B^0$ by
$B^0\ket{\psi_i}=\alpha_{i+1}\ket{\psi_i}$. Therefor the quantum
components $A_{k}^\varepsilon, \varepsilon\in\{+,-,0\}$, of
adjacency matrix $A_k$ is holds that
\begin{equation}
\lim_{k\longrightarrow\infty}\frac{A_{k}^{\varepsilon}}{\sqrt{p_{11}^{0}(k)}}=\lim_{k\longrightarrow\infty}\frac{A_{k}^{\varepsilon}}{\sqrt{k}}=B^{\varepsilon}
\;\;\;\ \varepsilon\in\{+,-,0\},
\end{equation}
in the stochastic sense. Then we have
\begin{equation}
\lim_{k\longrightarrow\infty}\braket{\phi_m}{\frac{A_{k}^{\varepsilon}}{\sqrt{k}}|\phi_0}=\braket{\psi_m}{B^{\varepsilon}|\psi_0},
\end{equation}
for $\varepsilon\in\{+,-,0\}$.

 To state a quantum central limit theorem for CTQW on odd graph
$O_k$, it is convenient to calculate amplitudes of probability as
\begin{equation}
q_m(t)=\lim_{k\longrightarrow
\infty}\braket{\phi_m}{e^{\frac{-itA_k}{\sqrt{k}}}|\phi_0}=\frac{1}{\sqrt{\omega_1\omega_2...\omega_m}}\int_{R}e^{-ixt}P_{m}(x)\mu_{\infty}(dx).
\end{equation}
According to the above theorem, we need only to find  two Szeg\"{o}-
Jacobi sequences $\{\omega_i\}$ and $\{\alpha_i\}$, where by using
Eqs.(\ref{central1}) and (\ref{central2}) we obtain as follows:\\
 if $i$ is odd,
\begin{equation}
\omega_i=\lim_{k\longrightarrow \infty}\frac{1}{k}
\frac{i+1}{2}(k-\frac{i-1}{2})=\frac{i+1}{2},
\end{equation}
if $i$ is  even,
\begin{equation}
\omega_i=\lim_{k\longrightarrow \infty}\frac{1}{k}
\frac{i}{2}(k-\frac{i}{2})=\frac{i}{2},
\end{equation}
if $i$ is odd or even,
\begin{equation}
\alpha_i=0.
\end{equation}
Thus, $\{\omega_i\}=\{1,1,2,2,3,3,4,4,...\}$ as desired. Therefore
Stieltjes transform  of infinite odd graphs is
\begin{equation}\label{measurein}
G_{\mu_\infty}(z)=\frac{1}{z-\frac{1}{z-\frac{1}
{z-\frac{2}{z-\frac{2}{z-\frac{3}{z-\frac{3}{z-\cdots}}}}}}},
\end{equation}
where the spectral distribution $\mu_\infty(x)$ in the Stieltjes
transform is given by \cite{nob1}
\begin{equation}
\mu_\infty(x)=|x|e^{-x^2}.
\end{equation}
Therefore we obtain the amplitude of probability at time $t$ and the
$0$-th stratum (starting vertex)
$$
q_0(t)=\int_{R}e^{-ixt}\mu_{\infty}(dx)=
\int_{-\infty}^{\infty}e^{-ixt}|x|e^{-x^2}dx
$$
\begin{equation}
=\frac{i\sqrt{\pi}t}{2}erf(it/2)e^{-\frac{t^2}{4}}+1.
\end{equation} In the above calculation we used formulas:
$$
x^ne^{-ixt}=i^n \frac{d^n}{d t^n}e^{-ixt},
$$
\begin{equation}\label{deriv1}
\int_{0}^{\infty}e^{-x^2}e^{-ixt}dx=\frac{\sqrt{\pi}}{2}(1-erf(it/2))e^{\frac{-t^2}{4}},
\end{equation}
where $erf(x)$  stands for the error function and it is defined as:
\begin{equation}
erf(x)=\frac{2}{\sqrt{\pi}}\int_{0}^{x}e^{-s^2}ds
\end{equation}
also, the derivative of the error function follows immediately from
its definition $\frac{\partial}{\partial
x}erf(x)=\frac{2}{\sqrt{\pi}}e^{-x^2}$.

By using Eqs.(\ref{v4}) and (\ref{deriv1}), we obtain the amplitude
of probability for walk at time $t$ and $m$-th stratum on infinite
odd
graphs  in terms of $q_0(t)$ as:\\
if $m$ is odd
\begin{equation}
q_{m}(t)=\frac{1}{(\frac{m-1}{2})!(\frac{m+1}{2})!}P_m(i\frac{d}{dt})q_0(t),
\end{equation}
if $m$ is even
\begin{equation}
q_{m}(t)=\frac{1}{((\frac{m}{2})!)^2}P_m(i\frac{d}{dt})q_0(t),
\end{equation}
where polynomials  $\{P_m(i\frac{d}{dt})\}$ are defined recurrently
by relation (\ref{op}) in terms of $i\frac{d}{dt}$. In the end for
example we obtain the amplitude of probability for walk at stratum
$1,2$ and $3$ as follows:\\
$m=1$
\begin{equation}
q_1(t)=P_1(i\frac{d}{dt})q_0(t)=i\frac{d}{dt}q_0(t)=\frac{\sqrt{\pi}}{4}(t^2-2)erf(it/2)e^{\frac{-t^2}{4}}-it/2,
\end{equation}
$m=2$
$$
q_2(t)=P_2(i\frac{d}{dt})q_0(t)=((i\frac{d}{dt})^2-1)q_0(t)
$$
\begin{equation}
=\frac{1}{8}(-i\sqrt{\pi}t^3\;erf(it/2)e^{\frac{-t^2}{4}}-2t^2+2i\sqrt{\pi}t\;erf(it/2)e^{\frac{-t^2}{4}}),
\end{equation}
$m=3$
$$
q_3(t)=\frac{1}{\sqrt{2}}P_3(i\frac{d}{dt})q_0(t)=\frac{1}{\sqrt{2}}((i\frac{d}{dt})^3-2i\frac{d}{dt})q_0(t)
$$
\begin{equation}
=\frac{1}{16\sqrt{2}}(-\sqrt{\pi}t^4erf(it/2)e^{\frac{-t^2}{4}}+2it^3+4\sqrt{\pi}t^2erf(it/2)e^{\frac{-t^2}{4}}-4it+
4\sqrt{\pi}erf(it/2)e^{\frac{-t^2}{4}}).
\end{equation}
\section{Conclusion}
In this paper by using the method of calculation of the probability
amplitude for continuous-time quantum walk on graph via quantum
probability theory, we have studied continuous-time quantum walk on
odd graph when the odd graph grow as time goes by. We have discussed
this question as a quantum central limit theorem for CTQW. It is
interesting to investigate CTQW on a growing family of graphs
 since it is probable the probability amplitudes of
CTQW to converge the uniform distribution, which is under
investigation.

\vspace{1cm} \setcounter{section}{0}
 \setcounter{equation}{0}
 \renewcommand{\theequation}{A-\roman{equation}}
  {\Large{Appendix A}}\\
\textbf{\large{Determination of spectral distribution by continued
fractions method }}

In this appendix we explain how we can determine spectral
distribution $\mu(x)$ of the graphs, by using the Szeg\"{o}-Jacobi
sequences $(\{\omega_k\},\{\alpha_k\})$, which the parameters
$\omega_k$ and $\alpha_k$ are defined in the subsection 3.2.

To this aim we may apply the canonical isomorphism from the
interacting Fock space onto the closed linear span of the orthogonal
polynomials determined by the Szeg\"{o}-Jacobi sequences
$(\{\omega_i\},\{\alpha_i\})$. More precisely, the spectral
distribution $\mu$ under question is characterized by the property
of orthogonalizing the polynomials $\{P_n\}$ defined recurrently by
$$ P_0(x)=1, \;\;\;\;\;\
P_1(x)=x-\alpha_1,$$
\begin{equation}\label{op}
xP_n(x)=P_{n+1}(x)+\alpha_{n+1}P_n(x)+\omega_nP_{n-1}(x),
\end{equation}
for $n\geq 1$.

As it is shown in \cite{tsc}, the spectral distribution ì can be
determined by the following identity:
\begin{equation}\label{v3}
G_{\mu}(z)=\int_{R}\frac{\mu(dx)}{z-x}=\frac{1}{z-\alpha_1-\frac{\omega_1}{z-\alpha_2-\frac{\omega_2}
{z-\alpha_3-\frac{\omega_3}{z-\alpha_4-\cdots}}}}=\frac{Q_{n-1}^{(1)}(z)}{P_{n}(z)}=\sum_{l=1}^{n}
\frac{A_l}{z-x_l},
\end{equation}
where $G_{\mu}(z)$ is called the Stieltjes transform and $A_l$ is
the coefficient in the Gauss quadrature formula corresponding to the
roots $x_l$ of polynomial $P_{n}(x)$ and where the polynomials
$\{Q_{n}^{(1)}\}$ are defined
recurrently as\\
        $Q_{0}^{(1)}(x)=1$,\\
    $Q_{1}^{(1)}(x)=x-\alpha_2$,\\
    $xQ_{n}^{(1)}(x)=Q_{n+1}^{(1)}(x)+\alpha_{n+2}Q_{n}^{(1)}(x)+\omega_{n+1}Q_{n-1}^{(1)}(x)$,\\
    for $n\geq 1$.

Now if $G_{\mu}(z)$ is known, then the spectral distribution $\mu$
can be recovered from $G_{\mu}(z)$ by means of the Stieltjes
inversion formula:
\begin{equation}\label{m1}
\mu(y)-\mu(x)=-\frac{1}{\pi}\lim_{v\longrightarrow
0^+}\int_{x}^{y}Im\{G_{\mu}(u+iv)\}du.
\end{equation}
Substituting the right hand side of (\ref{v3}) in (\ref{m1}), the
spectral distribution can be determined in terms of $x_l,
l=1,2,...$, the roots of the polynomial $P_n(x)$, and  Guass
quadrature constant $A_l, l=1,2,... $ as
\begin{equation}\label{m}
\mu=\sum_l A_l\delta(x-x_l)
\end{equation}
 ( for more details see Refs. \cite{js,jsa,tsc,st}.)

\newpage
{\bf Figure Captions}

{\bf Figure-1:} The Petersen graph.

\end{document}